\documentclass[12pt]{article}

\usepackage{sbc-template}
\usepackage{xcolor}
\usepackage{verbatim}
\usepackage{graphicx,url}
\usepackage[utf8]{inputenc}
\usepackage[english]{babel}
\usepackage{subcaption}
\usepackage{booktabs}

\title{Do Electromagnetic Side-Channel Attacks Threaten
Electronic Polling Stations? Scenarios and Recommendations}

\author{Lucas Brito\inst{1},
Leonardo Teodoro\inst{1},
Pedro Tomaz\inst{1},
Alyson Isaluski\inst{1},
Leandro Hyeda\inst{1},\\
Antonio Oliveira-Jr\inst{2,}\inst{3},
Saulo Queiroz\inst{1}
}
\address{
Federal University of Technology -- Paraná (UTFPR)\\
Ponta Grossa -- PR -- Brazil
\email{\{lbrito,
lteodoro,
pedrotomaz,}
\email{
alysonisaluski,
leandrohyeda\}@alunos.utfpr.edu.br},
\email{antoniojr@ufg.br, sauloqueiroz@utfpr.edu.br}
\nextinstitute
Federal University of Goiás (UFG)\\
Goiânia -- GO -- Brazil.
\nextinstitute
Fraunhofer Portugal AICOS, Porto 4200-135, Portugal.%
}
\begin{document}

\maketitle

\begin{abstract}
This paper investigates the threat to ballot secrecy in the Brazilian
 electronic voting machine (UEB) posed by electromagnetic side-channel
attacks, also known as TEMPEST attacks. In these attacks, screen content
can be reconstructed remotely by intercepting electromagnetic emanations
associated with the target device's video signal. This work is motivated
by a recent ruling by a Brazilian electoral court concerning an attempt
to violate ballot secrecy using electronic equipment. Based on publicly
available information about the electoral system, attack scenarios
against polling stations are proposed. Experiments using software-defined
radio show that the effectiveness of TEMPEST attacks strongly depends on
the lack of oversight resulting from public unawareness of the threat.
Finally, awareness guidelines are proposed for voters, poll workers, and
party representatives to mitigate attack risks within a polling station.
\end{abstract}

\section{Introduction}
Recently, the Regional Electoral Court of Pará (TRE-PA) removed a city
councilor in Ourilândia do Norte, Pará, from office for recruiting voters
to wear glasses with embedded miniature cameras to violate ballot secrecy
in the Brazilian electronic voting machine (UEB)~\cite{tre_pa_cassacao}.
In a related context, this work investigates threats to ballot secrecy
from electromagnetic side-channel attacks, also known as TEMPEST
attacks~\cite{tempestclass-ieeeaccess2025}.

In a TEMPEST attack against a graphics system, an adversary uses a
software-defined radio (SDR) to intercept the electromagnetic emanations
inherent in the transmission of the video signal through the system's
components (e.g., cables, connectors, circuits, and display). With
appropriate processing of the captured signal, a successful attack can
reconstruct the video image carried by that signal. This attack
compromises only the system's confidentiality, leaving its integrity and
availability unaffected. Since the image must be displayed in a form
intelligible to humans, traditional data encryption methods do not
directly mitigate this type of threat.

The first demonstration of a TEMPEST attack reported in the academic
literature took place in the 1980s and became known as ``van Eck
\emph{phreaking}'' after its author~\cite{vaneck1985tempest}. At the time,
increasing display resolutions posed natural obstacles to exploiting the
threat, since both the computational complexity and the required sampling
rate were proportional to the target display's pixel rate. Furthermore,
commercial SDRs capable of meeting the attack's acquisition and
processing requirements were unavailable. Once these technological
limitations were overcome, TEMPEST attacks began to be investigated
using different techniques, scenarios, and devices, including
multi-display systems~\cite{choi2024analysis},
smartphones~\cite{cheng2025capacitive}, and deep learning
approaches~\cite{deepl-larroca-2024}.

Regarding electronic voting systems, \cite{riccps1634} refers to a
TEMPEST experiment based on acoustic side channels involving the UEB's
keypad, whereas~\cite{rohr2009urna} mentions evidence of electromagnetic
eavesdropping involving a voting machine model different from the
Brazilian one. Both reports lack sufficient technical detail to ensure
experimental reproducibility. In~\cite{publicTEMPEST2026}, the authors
investigate spectral signatures inherent in TEMPEST signals associated
with the UEB's graphical interface, but do not consider operational
attack scenarios or discuss potential vulnerabilities inherent in a
Brazilian polling station, as addressed in this study.

\section{Video Signal}\label{sec:background}
The VGA signal, used by at least two UEB models in
operation\footnote{\url{https://www.tre-sc.jus.br/eleicoes/urna-eletronica/modelos/}.},
employs pulse amplitude modulation (PAM), in which pixel information
modulates the amplitude of time-shifted rectangular pulses. In this
modulation, each discrete pixel $x[n]$ modulates the amplitude of a
rectangular pulse $p(t)$ of duration $T_p$, producing a continuous-time
signal modeled as

\begin{equation}
x(t) = \sum_{n=-\infty}^{\infty} x[n]p(t-n T_p),
\label{eqn:sinaltempest}
\end{equation}

where $x(t)$ represents the continuous-time signal transmitted at time
$t$, and $T_p$ is the time interval associated with each pixel, given by
\begin{equation}
T_p = \frac{1}{P_x P_y f_v}
\quad \textrm{seconds},
\label{eqn:tp}
\end{equation}

where $P_x$, $P_y$, and $f_v$ denote the total number of pixels per line
in a display frame, the number of lines per frame, and the frame refresh
rate, respectively. The corresponding pixel rate is $f_p=1/T_p$
(commonly expressed in hertz, Hz). Note that $P_x$ and $P_y$ also include
so-called \emph{blank pixels}, which are used for synchronization between
the graphics card and the monitor and do not represent visible content.
The number of these pixels for each resolution is defined by the VESA
standard~\cite{vesa_dmt_2013}.

Applying the Fourier transform to the signal in
(\ref{eqn:sinaltempest}) yields

\begin{equation}
X(f)=\mathcal{F}\{x(t)\}
=
P(f)\sum_n x[n]e^{-j2\pi f nT_p},
\label{eqn:ft}
\end{equation}

where

\begin{equation}
P(f)=\mathcal{F}\{p(t)\}=T_psinc(fT_p)
\end{equation}

is the Fourier transform of the rectangular pulse, while the summation
term corresponds to the discrete-time Fourier transform (DTFT) of the
pixel sequence. The signal spectrum contains spectral components spaced
by the pixel rate $\Delta f = f_p = 1/T_p$ Hz and modulated by the
spectral envelope $P(f)$. Although the $sinc$ function has zeros exactly
at integer multiples of $f_p$---i.e., frequencies at which an SDR may
successfully perform a TEMPEST attack---in practice, bandwidth
limitations, noninstantaneous pulse rise times, and distortions
introduced by the electromagnetic channel alter the ideal rectangular
pulse shape. Consequently, the TEMPEST signal still exhibits observable
energy near these frequencies, allowing an adversary to exploit these
spectral components to reconstruct the image. Further technical details
on TEMPEST attacks can be found
in~\cite{lvds-ieeetifs-25,tempestclass-ieeeaccess2025,grtempest-2022}.
\begin{figure}[ht]
    \centering
        \centering
        \includegraphics[width=7cm]{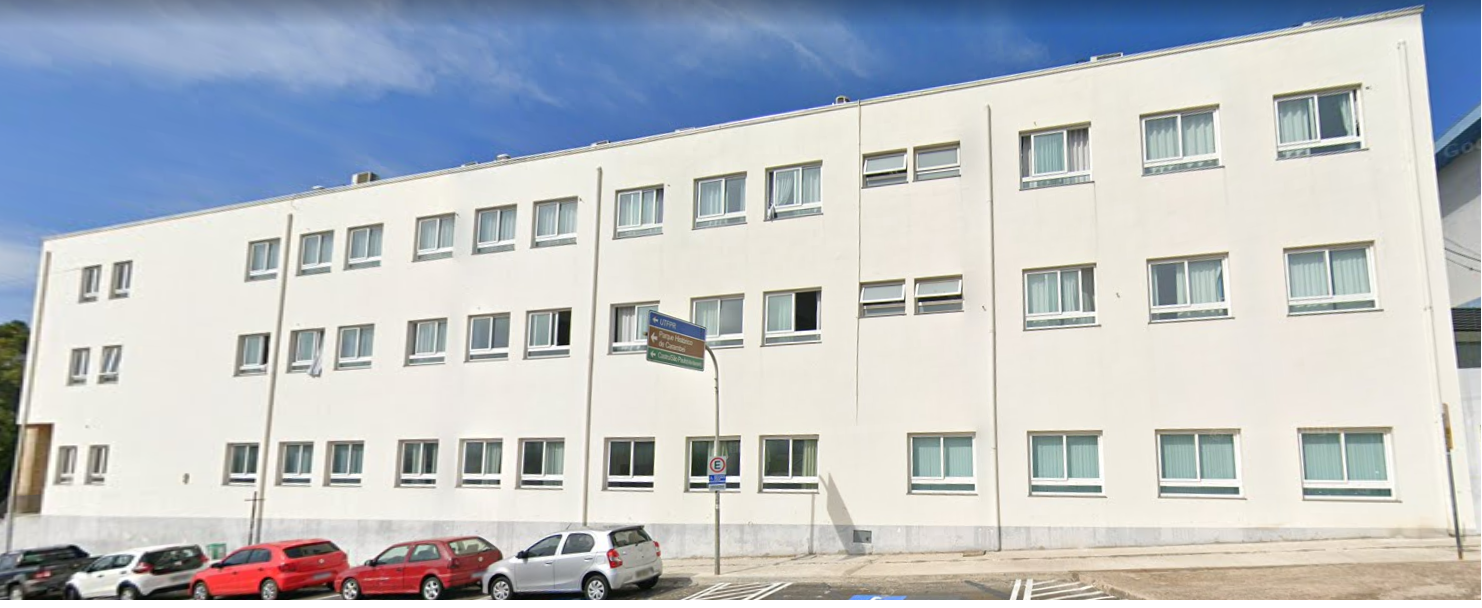}
        \caption{Polling station consistent with the proposed attack scenarios.}
        \label{fig:secao1}
\end{figure}

\section{Attack Scenarios and Experiments}\label{sec:tse}
According to current information from Brazil's Superior Electoral Court
(TSE), the Brazilian electronic voting system uses the UE2013, UE2015,
UE2020, and UE2022 voting machine models. Each machine is installed in an
individual voting booth inside a polling station. During voting hours,
various participants have physical access to the polling station,
including voters, poll workers, and party observers. Furthermore, polling
stations are frequently set up in buildings, preferably public ones,
temporarily made available to the TSE for use as polling places. Based on
these characteristics, we identify two potential TEMPEST attack
scenarios against a Brazilian polling station:

\begin{itemize}
\item \textbf{Scenario 1:} In this scenario, an adversary with access to
a space adjacent to the polling station room (e.g., another room, a
corridor, or a public street) installs an SDR--computer setup that can
operate locally or remotely if the computer is connected to the
Internet. The attack exploits the fact that voting booths are often
positioned near walls to prevent others from seeing a voter's choices.
Under these conditions, physical proximity may facilitate the capture
of electromagnetic leakage.

\item \textbf{Scenario 2:} In this scenario, a person with legitimate
access to the polling station room may be recruited to carry a portable,
battery-powered SDR capable of storing raw signals for subsequent
\emph{offline} analysis during voting hours. Examples of SDRs suitable
for this scenario include the Ettus USRP E312 and the USRP-LW E3xx. If
properly executed, this attack may circumvent the minimum distance of
0.5 meters beyond which the electromagnetic attenuation expected in the
UE2020 and UE2022 models renders a TEMPEST attack
infeasible~\cite{tse_ue2020_security_specs}.
\end{itemize}
Figure~\ref{fig:secao1} shows polling station No.~47 of the 14th electoral
zone in Ponta Grossa, Paraná, as a real-world example of a station
consistent with Scenarios 1 and 2 described in this work. The image shows
rooms adjacent to one another and to the public street, indicating that
a voting booth may be separated from these spaces by only a masonry
wall.

All tests used resolutions of $1280 \times 768@60$ (i.e., $f_p=74.25$
MHz), consistent with the UE2013 and UE2015
models\footnote{\url{https://www.justicaeleitoral.jus.br/tps/arquivos/2021/apresentacao-7-tps-2021-cotel-urna-eletronica.pdf}.},
and $1920 \times 1080@60$ (i.e., $f_p=148.5$ MHz, \emph{Full HD}), which
exceeds the minimum resolution required for the UE2020
models~\cite{tse_ue2020_security_specs_resolucao}. To the best of the
authors' knowledge, no public documents provide information on the
UE2022 model's resolution. In such cases, an estimate can be obtained
using discrete autocorrelation algorithms based on the fast Fourier
transform (FFT) and related
approaches~\cite{disp-eleccompa-2019,sic-magazine-2025}.

In Scenario 1, the SDR was positioned approximately 1 meter from the UEB,
with a masonry wall separating them. In Scenario 2, the distance between
the SDR and the UEB was less than 0.5 meters, a condition potentially
favorable to TEMPEST attacks against the UE2020 and UE2022 models.
Video signals were intercepted using an Ettus USRP B200 radio connected
to a digital TV antenna and tuned to a harmonic of the pixel rate in
each experiment. The SDR samples were processed by a laptop with 16 GB
of RAM and a 12th-generation Intel i7 processor. Note that, in the
practical experiments, Scenario 2 was represented by the physical
proximity between the SDR and the target monitor, since the SDR model
used does not support internal sample storage for subsequent processing.
Across different experiments, visually legible signals were identified
at the 3rd, 6th, and 7th harmonics using an SDR sampling rate of
$f_s = 54$ MHz. The collected signals were subsequently resynchronized
based on the known resolutions and converted into images using the
GNU Radio gr-tempest module~\cite{grtempest-2022}.

\begin{figure}[ht]
    \centering
    \begin{subfigure}{0.48\linewidth}
        \centering
        \includegraphics[width=0.83\linewidth]{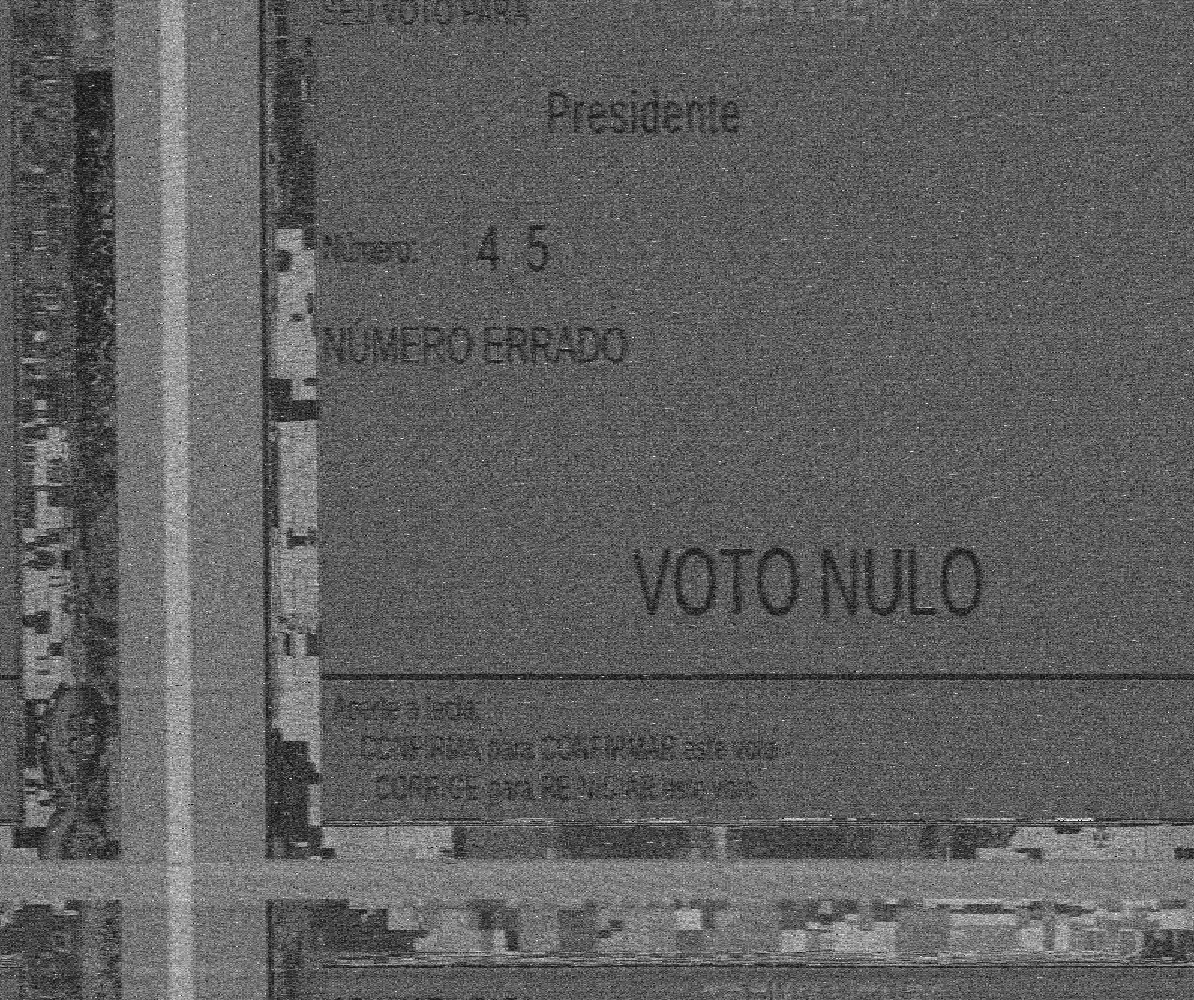}
        \caption{Scenario 1: $\approx$ 1 meter, with a wall.}
        \label{fig:cen1}
    \end{subfigure}
    \hfill
    \begin{subfigure}{0.48\linewidth}
        \centering
        \includegraphics[width=0.83\linewidth]{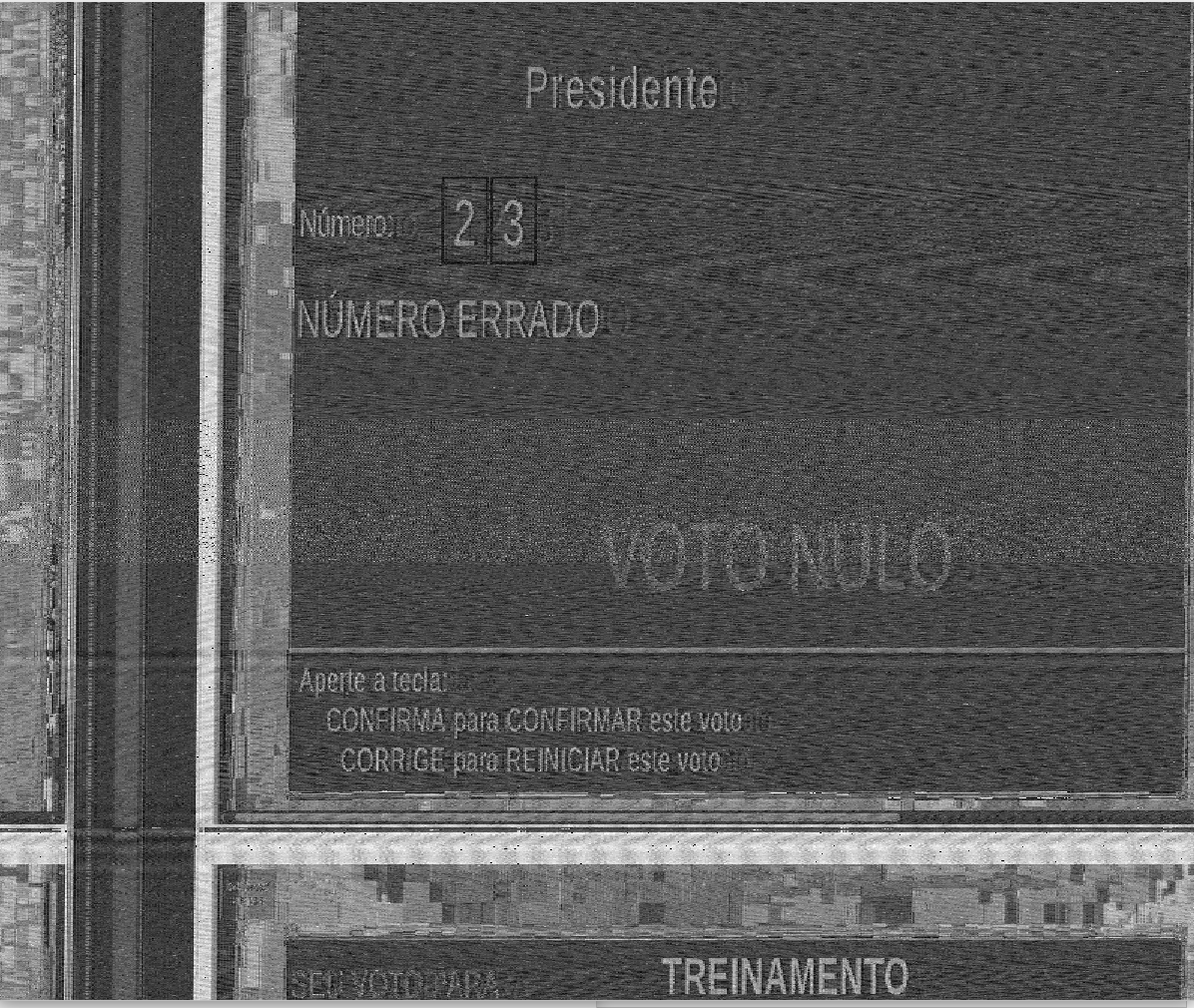}
        \caption{Scenario 2: less than 0.5 meters, without obstacles.}
        \label{fig:cen2}
    \end{subfigure}
   \caption{Images reconstructed through a TEMPEST attack across a masonry wall (left) and without obstacles (right).}
    \label{fig:geral}
\end{figure}

As shown in Figs.~\ref{fig:cen1} and \ref{fig:cen2}, the candidate number
displayed on the respective target monitor could be identified in both
experimental scenarios. Similar results were also obtained by varying
the UEB images, the signal type (VGA/HDMI), and the resolution. Owing to
space limitations, a summary of the results and the artifacts required
to reproduce the experiments are available in the lead author's
repository\footnote{{\url{https://github.com/kewlzin/sbseg-ueb}.}}.
Additional scenarios with more intervening walls and greater distances
between the SDR and the target monitor were also investigated. In these
cases, visually intelligible images could not be reconstructed because
of the severe distortion of the captured signal. Thus, the results
indicate that the risk of violating ballot secrecy in the UEB through
electromagnetic TEMPEST attacks strongly depends on the physical
proximity between the adversary and the polling station. We therefore
recommend informing poll workers, voters, and party representatives
about both the existence of TEMPEST attacks and the typical appearance
and operating characteristics of SDR devices to strengthen oversight
and deter this type of threat.

\section{Conclusions and Recommendations}\label{sec:conclusao}

This work investigated the possibility of violating ballot secrecy in
the Brazilian electronic voting machine (UEB) through electromagnetic
side-channel attacks using software-defined radios (SDRs). Two attack
scenarios were proposed and experimentally evaluated using a VGA monitor
reproducing the visual characteristics of the UEB interface. The first
scenario considered an adversary with access to a space adjacent to the
polling station room (e.g., another room, a corridor, or a public
street), located approximately 1 meter from the UEB and separated from
it by a masonry wall. The second scenario assumed that a person with
legitimate access to the polling station (e.g., a party observer, voter,
or poll worker) had been recruited to carry a compact, battery-powered
radio device capable of storing signals for subsequent \emph{offline}
analysis.

Information displayed on the target monitor could be inferred in all
scenarios considered. This was not the case at greater distances and/or
in the presence of additional obstacles. Overall, we conclude that the
risk associated with TEMPEST attacks against the UEB strongly depends on
the physical proximity between the adversary and the polling station,
highlighting the importance of awareness efforts aimed at strengthening
oversight. Specifically for the first scenario, we recommend avoiding
placement of the UEB next to walls adjoining spaces subject to limited
oversight or monitoring. Regarding the second scenario, we conclude that
the risk of violating ballot secrecy using SDRs has operational
similarities to cases involving other portable electronic devices, such
as the incident involving \emph{smart glasses} that led to a city
councilor's removal from office in Ourilândia do Norte, Pará. In this
context, public unawareness of SDRs and the threat posed by TEMPEST
attacks may allow an adversary to operate without arousing suspicion.
We therefore recommend informing poll workers, voters, and party
representatives about both the existence of TEMPEST attacks and the
typical appearance and operating characteristics of SDR devices.
Finally, we believe this work may encourage further discussion of
electromagnetic security, public transparency, and the mitigation of
side-channel attacks in the Brazilian electoral context.

\section*{Acknowledgments}
This work was partially funded by the \emph{Advanced Multimodal Sensing}
(AIMS) project, supported by the \emph{Advanced Knowledge Center in
Immersive Technologies} (AKCIT), with funding from MCTI's PPI IoT program
under Agreement No.~057/2023 with EMBRAPII. The authors also thank the
Goiás Research Foundation (FAPEG) for the financial support provided for
this research (Project No.~64448878/2024).
\bibliographystyle{sbc}
\bibliography{refs}

\end{document}